\documentstyle[prl,aps,multicol,epsf,rotate]{revtex}
\begin{document}
\draft
\preprint{XXXX}
\title{Directed Percolation and Generalized Friendly Walkers}
\author{John Cardy$^{1,2}$ and Francesca Colaiori$^{1}$}
\address{$^1$University of Oxford, Department of Physics -- Theoretical
         Physics, 1 Keble Road, Oxford OX1 3NP, U.K. \\
         $^2$All Souls College, Oxford.}
%
%
\maketitle
\begin{abstract}
We show that the problem of directed percolation on an arbitrary
lattice is equivalent to the problem of $m$ directed random walkers
with rather general attractive interactions, when suitably continued
to $m=0$. In 1+1 dimensions, this is dual to a model of interacting steps
on a vicinal surface. A similar correspondence with interacting
self-avoiding walks is constructed for isotropic percolation.
\end{abstract}
\begin{multicols}{2}
The problem of directed percolation (DP), first introduced by Broadbent and
Hammersley \cite{Broadbent}, continues to attract interest even though it
has so far defied all attempts at an exact solution, even in two
dimensions. Although the problem was
originally formulated statically on a lattice with a preferred
direction, when the latter is interpreted as time the
universal behavior close to the percolation threshold is
also believed to describe the transition from a noiseless absorbing
state to a noisy, active one, which occurs in a wide class of
stochastic processes \cite{Dickman}. 
It also maps onto reggeon field theory, which
describes high-energy diffraction scattering in particle physics
\cite{CS}.

\begin{figure}
\centerline{
\epsfxsize=3in
\epsfbox{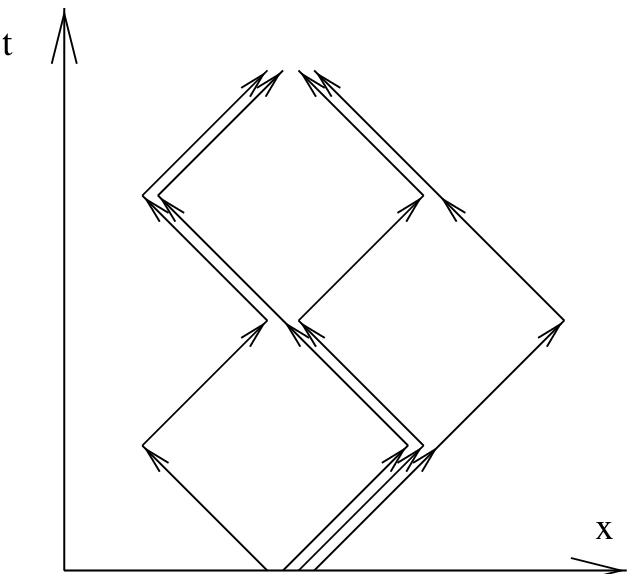}}
\caption{Typical configuration of $m=4$ friendly walkers on the diagonal
square lattice.}
\label{fig1}
\end{figure}

Some time ago, Arrowsmith, Mason and Essam \cite{AME} argued that
the pair connectedness probability $G(r,r')$ for
directed bond percolation on a two-dimensional diagonal square lattice
can be related to the partition function for the weighted paths of
$m$ `friendly' 
walkers which all begin at $r$ and end at $r'$, when suitably continued 
to $m=0$. These are directed random walks which may share bonds of the
lattice but do not cross each other (see Fig.~\ref{fig1}). In fact,
Arrowsmith et al \cite{AME} represented these configurations
in other ways: either as
\em vicious \em walkers, which never intersect, by moving the friendly
walkers each one
lattice spacing apart horizontally; or as \em integer flows \em on  the
directed lattice, to be defined explicitly below. Arrowsmith and Essam
\cite{AE} showed that $G(r,r')$ is also related to a partition
function for a $\lambda$-state chiral Potts model on the dual lattice,
on setting $\lambda=1$, thus generalizing the well-known result of
Fortuin and Kasteleyn\cite{KF} for ordinary percolation.

In a more recent paper, Tsuchiya and Katori \cite{TK} have considered instead
the order parameter of the DP problem, and have shown that 
in d=2 
it is related
to a certain partition function of the same $\lambda=1$ chiral Potts
model, and also that, for arbitrary $\lambda$, the latter is equivalent
to a partition function for $m=(\lambda-1)/2$ friendly walkers. 

It is the purpose of this Letter to describe a broad generalization of
these results. We demonstrate, in particular, a direct connection between a
general connectedness function of DP and a corresponding partition
function for $m$ friendly walkers when continued to $m=0$. We show that
this holds on an arbitrary directed lattice in any number of dimensions,
and for all variant models of DP, whether bond, site or correlated.
Moreover the weights for a given number of walkers passing along a given
bond or through a given site may be chosen in a remarkably arbitrary
fashion, still yielding the same result at $m=0$. 

We now describe the correspondence between these two problems in detail.
A \em directed lattice \em is composed of
a set of points in ${\bf R}^d$ with a privileged coordinate $t$,
which we may think of as time.
Pairs of these sites $(r_i,r_j)$ are connected by fixed bonds, oriented
in the direction of increasing $t$, to form a directed lattice. In the
directed \em bond \em problem, each bond is  open with a
probability $p$ and closed with a probability $1-p$, and in the \em site
\em problem it is the sites which have this property. In principle the
probabilities $p$ could be inhomogeneous, and we could also consider
site-bond percolation and situations in which different bonds and sites
are correlated. Our general result applies to all these cases, but for
clarity we shall restrict the argument to independent homogeneous
directed bond percolation. The pair connectedness $G(r,r')$ is the
probability that the points $r$ and $r'$ (with $t<t'$) are connected by
a continuous path of bonds, always following the direction of increasing
$t$. 

On the same lattice, let us define the corresponding \em integer flow
\em problem. Assign a non-negative integer-valued current
$n(r_i,r_j)$ to each bond, in
such a way that it always flows in the direction of increasing $t$,
and is conserved at the vertices. At the point $r$ there is a source
of strength $m\geq1$, and at $r'$ a sink of the same strength. There
is no flow at times earlier than that of $r$ or later than that of $r'$.
Such a configuration may be thought of as representing the worldlines of
$m$ particles, or walkers, where more than one walker may share the same
bond. The configurations are labeled by distinct allowed values of
the $n(r_i,r_j)$, so that they are counted in the same
way as are those of identical bosons. Alternatively, in $1+1$ dimensions, we
may regard the walkers as distinct but with worldlines which are
not allowed to cross.
In the partition sum, each bond is counted with a weight
$p(n(r_i,r_j))$. In the simplest case we take $p(0)=1$ and $p(n)=p$ for
$n\geq1$ (although we shall show later that this may be generalized).
Since $p>p^n$ for $n>1$, there is an effective attraction
between the walkers, leading to the description `friendly'.
The partition function is then
\begin{displaymath}
Z(r;r';m)\equiv\sum_{\rm allowed\ configs}\,\prod_{(r_i,r_j)}p(n(r_i,r_j))
\end{displaymath}
This expression is a polynomial in $m$ and so may be
evaluated at $m=0$. The statement of the correspondence between DP and
the integer flow problem for the case of the pair connectedness is then
\begin{displaymath}
G(r;r')=Z(r;r';0).
\end{displaymath}
Note that since the weights $p(n)$ behave non-uniformly as $n\to0$, the
continuation of $Z(r;r';m)$ to $m=0$ is not simply the result of taking
zero walkers (which would be $Z=1$): rather it is the non-trivial answer
$G$.
Similar results hold for more generalized connectivities. For example,
if we have points $(r'_1,r'_2,\ldots,r'_l)$ all at the same time $t'>t$, we
may consider the probability $G(r;r'_1,r'_2,\ldots,r'_l)$
that all these points, irrespective of any others,  
are connected to $r$. 
The corresponding integer
flow problem has a source of strength $m\geq l$ at $r$, and sinks of arbitrary
(but non-zero) strength at each point $r'_j$. In this case
\begin{displaymath}
G(r;r'_1,r'_2,\ldots,r'_l)=(-1)^{l-1}Z(r;r'_1,r'_2,\ldots,r'_l;m=0)
\end{displaymath}
where the partition function is defined with the same weights as before.
Since the order parameter for DP may be defined as the limit as
$t'-t\to\infty$ of $P(t'-t)$, the probability that \em any \em site at
time $t'$ is connected to $r$, and this may be written using an
inclusion-exclusion argument as 
\begin{displaymath}
P(t'-t)=\sum_{r'}G(r;r')-\sum_{r'_1,r'_2}G(r;r'_1,r'_2)+\cdots
\end{displaymath}
(where the sums over the $r'_j$ are all restricted to the fixed time
$t'$),
we see that it is in fact given by the $m=0$ evaluation of the partition
function for \em all \em configurations of $m$ walkers which begin at $r$ and
end at time $t'$. This generalizes the result of Tsuchiya and Katori
\cite{TK} to an arbitrary lattice.
Although this continuation to $m=0$ is reminiscent
of the replica trick, it is in fact quite different. Moreover it is
mathematically well-defined, since, as we argue below, $Z$ is a finite
sum of terms, each of which, with the simple weights given above, is a
polynomial in $m$.

\begin{figure}
\centerline{
\epsfxsize=3.2in
\epsfbox{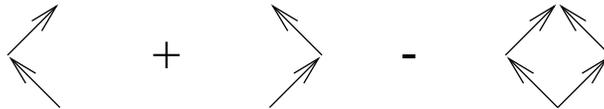}}
\caption{Allowed graphs $\cal G$ corresponding to $G((0,0);(0,2))$. The
first two are counted with weight $p^2$. The last is necessary to avoid
double-counting, and comes with weight $-p^4$.}
\label{fig2}
\end{figure}

\begin{figure}
\centerline{
\epsfxsize=3.2in
\epsfbox{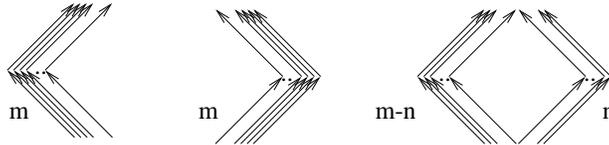}}
\caption{Sets of configurations of $m$ friendly walkers corresponding to each of
the graphs in Fig.~2. The last corresponds to $m-1$ configurations.}
\label{fig3}
\end{figure}

We now give a summary of the proof, which is elementary. 
The connectedness function $G(r;r'_1,r'_2,\ldots,r'_l)$
is given \cite{Essam}
by the weighted sum of all graphs $\cal G$ which have the property
that each vertex may be connected backwards to $r$ and forwards to at
least one of the $r'_j$. (Alternatively, $\cal G$ is a union of directed
paths from $r$ to one of the $r'_j$.) 
Each such graph is weighted by a factor $p$ for
each bond and $(-1)$ for each closed loop. A simple example is shown in
Fig.~\ref{fig2}. A given graph corresponds to summing over
all configurations in which
the bonds in $\cal G$ are open, irrespective of all other bonds in the
lattice. The factors of $(-1)$ are needed to eliminate double-counting.
It is useful to decompose vertices in $\cal G$ 
with coordination number $>3$ by
inserting permanently open bonds into them in such a way that the 
only vertices are those
in which two directed bonds merge to form one ($2\to1$), and vice
versa. This does not affect the connectedness properties. We may
then associate the factors of $(-1)$ with each $1\to2$ vertex in $\cal
G$, as long as we incorporate an overall factor $(-1)^{l-1}$ in $G$.
With each graph $\cal G$ we associate a restricted set of integer
flows,
called proper flows,
such that $n\geq1$ for each bond in $\cal G$, and $n=0$ on each bond not
in $\cal G$. Those corresponding to the graphs in Fig.~\ref{fig2} are
shown in Fig.~\ref{fig3}. Note that the last graph corresponds to $m-1$
configurations of integer flows, which gives precisely the required
factor of $(-1)$ when we set $m=0$. In general, summing over all allowed
integer flows will generate the sum over all allowed $\cal G$,
with correct weights $p$:
the non-trivial part is to show that we recover the correct
factors of $(-1)$ when we set $m=0$.

This follows from the following simple lemma: if $A(n)$ is a polynomial
in $n$, and we define the polynomial $B(m)\equiv\sum_{n=1}^{m-1}A(n)$,
then $B(0)=-A(0)$. We give a proof which shows that the result may be
generalized to other functions: write $A(n)$ as a Laplace transform
$A(n)=\int_C(ds/2\pi i)e^{ns}\tilde A(s)$. Then
$B(m)=\int_C(ds/2\pi i)\big((e^s-e^{ms})/(1-e^s)\big)\tilde A(s)$,
so that $B(0)=-\int_C(ds/2\pi i)\tilde A(s)=-A(0)$. An immediate
corollary is that if $A(n_1,n_2,\ldots)$ is a polynomial in several
variables, and $B(m)\equiv\sum_{n=1}^{m-1}A(n,m-n,\ldots)$, then
$B(0)=-A(0,0,\ldots)$. We use this to proceed by induction on the
number of $1\to2$ vertices in $\cal G$. Beginning with the vertex which
occurs at the earliest time, the 
contribution to $Z$ from the proper flow on
$\cal G$, when evaluated at
$m=0$, is, apart from a factor $(-1)$, equal
to that for another graph $\cal G'$ which will have one fewer
$1\to2$ vertex. However, $\cal G'$ differs from the previously allowed
set of graphs $\cal G$ in that it may have more than one vertex at which
current may flow into the graph. For this reason we extend the definition
of the allowed set of graphs to include those in which every vertex is
connected to at least one `input' point
$(r_1,r_2,\ldots)$ and at least one `output' point
$(r'_1,r'_2,\ldots)$. In the corresponding integer flow problem,
currents $(m_1,m_2,\ldots)$ flow in at the inputs, whereas the only
restriction on the outputs is that non-zero current should flow out. The
partition function is then the weighted sum over all such allowed integer
flows. Induction on the number of $1\to2$ vertices then shows that this
partition function, evaluated at $m_j=0$, gives the corresponding DP
graph correctly weighted. (The induction starts from graphs with no 
$1\to2$ vertices which involve no summations and for which the result is
trivial.)

Since our main result relies only on the lemma
it follows also for rather general weights $p(n)$. The only
requirement is that $p(n)$ grow no faster than an exponential at large
$n$, and that, when continued to $n=0$, it give the value $p\not=1$.
In this case, $Z$ will no longer be a polynomial in $m$, but, since by
the inductive argument above it is given by a sum of convolutions of
$p(n)$, its continuation to $m=0$ will be well-defined through its
Laplace transform representation.
For example, we could take $p(n)=p^{1-n}$ for $n\geq1$.
This raises the possibility of choosing some suitable set of weights for
which the integer flow problem, at least in $1+1$ dimensions, is
integrable, for example by Bethe ansatz methods. Unfortunately our
results in this direction are, so far, negative.
In the case of bond percolation on a diagonal square lattice 
let $Z(x_1,x_2, \dots ,x_m;t)$ 
be the partition function under the constraint that the walkers arrive
at $\{x_1,x_2,\dots ,x_m\}$ at time $t$, the physical region being 
$\{ x_1 \leq x_2 \leq \dots \leq x_m \}$.
Turning the master equation for $Z$ in an eigenvalue problem 
\cite{note} and writing the eigenfunction $\psi_m(x_1,x_2,\dots x_m)$ 
in the usual Bethe ansatz form, one gets for 
$\psi_2(x_1,x_2)\!=\!A_{12}e^{i(x_1k_1+x_2k_2)}+
A_{21} e^{i(x_1k_2+x_2k_1)}$ the following condition
on the amplitudes: 
\begin{displaymath}
\frac{A_{21}}{A_{12}}=
-\frac{e^{\,\,i(k_1-k_2)}-\epsilon
\left(e^{i(k_1+k_2)}+e^{-i(k_1+k_2)}\right)}
{e^{\!-i(k_1-k_2)}-\epsilon
\left(e^{i(k_1+k_2)}+e^{-i(k_1+k_2)}\right)} 
\end{displaymath}
(the same as that which appears in the XXZ spin chain \cite{Yang}) 
where $\epsilon=p(2)/p(1)^2-1$. 
Requiring that the
$m$-particle scattering should factorise into a product of these
two-body $S$-matrices places constraints on the weights $p(n)$.
In general these equations appear too difficult to solve,
except in the weak interaction limit ($\epsilon\simeq 1/2$), 
where we find
\begin{displaymath}
2^n=2q(n)+\sum_{s=1}^{n-1} q(n-s) q(s) (1-\lambda s(n-s))+O(\lambda^2)
\end{displaymath}
where $q(s)=p(s)/(p(1))^s$, $\lambda=2\epsilon-1$. 
This may be solved for successive
$q(n)$, but it is easy to see by applying the above lemma that, when
continued to $n=0$, it will always yield the value $1$, rather than $p$
as required. We conclude that the $m=0$ continuation 
of this integrable case does not correspond
to DP. It is nevertheless interesting that integrable models of such
interacting walkers can be formulated.

In 1+1 dimensions, our generalized friendly walker model maps naturally onto
a model of a step of total height $m$ on a vicinal surface, by assigning
integer height variables $h(R)$ to the sites $R$ of the dual lattice, such
that $h=0$ for $x\to-\infty$, $h=m$ for $x\to+\infty$, and $h$ increases
by unity every time the path of a walker is crossed. The weights for
neighboring dual sites $R$ and $R'$ are $p(h(R')-h(R))$. This
is slightly different from, and simpler than, the
chiral Potts model studied in \cite{AE,TK}.

A similar correspondence between percolation and
interacting random walks is valid also for the
isotropic case. 
The pair connectedness $G(r,r')$ 
may be represented by a sum of graphs $\cal G$, just as in DP
\cite{Essam}. Each graph consists of a union of oriented paths 
from $r$ to $r'$, 
As before, each bond is counted with weight $p$ and
each loop carries a factor $(-1)$. Note that graphs which 
contain a closed loop of oriented bonds are excluded. Such
contributions cannot occur in DP because of the time-ordering.
The correspondence with integer flows or friendly walkers follows as
before. The latter picture is particularly simple. $m$ walkers begin at 
$r$ and end at $r'$. When two or more walkers occupy the same bond,
they must flow parallel to each other. 
Since they cannot form closed loops, they are
\em self-avoiding\em. Moreover, walkers other than those which
begin and end at $r$ and $r'$, which could also form closed loops, are
not allowed.  Each occupied bond has weight $p(n)$ as before, 
and the separate configurations are counted
using Bose statistics. $G(r,r')$ is then given by the continuation to
$m=0$ of the partition function. We conclude that ordinary percolation is
equivalent to the continuation to $m=0$ of
a problem of $m$ oriented self-avoiding
walks, with infinite repulsive interactions between anti-parallel
segments on the same bond, but attractive parallel interactions.
In two dimensions, this is again dual to an interesting height model,
in which neighboring heights satisfy $|h(R')-h(R)|\leq m$, but local
maxima or minima of $h(R)$ are excluded.
For example, the order parameter of percolation is
given by the continuation to $m=0$ of the partition function for 
a screw dislocation of strength $m$ in this model.

To summarize, we have shown that the DP problem is simply related to the
integer flow problem, or equivalently that of $m$ bosonic `friendly'
walkers, when suitably continued to $m=0$. This holds on an arbitrary
directed lattice in any number of dimensions, and with rather general
weights. It is to be hoped that this correspondence might provide a new
avenue of attack on the unsolved problem of directed percolation.

The authors acknowledge useful discussions with F.~Essler and
A.~J.~Guttmann, and
thank T.~Tsuchiya and M.~Katori for sending a copy of their paper prior
to publication. This research was supported in part by the Engineering and 
Physical Sciences Research Council under Grant GR/J78327.

\end{multicols}

\newpage


\begin{references}
\bibitem{note} We note that is necessary to make further assumptions 
in concluding that the \em eigenvalues \em of the DP transfer matrix
correspond to the continuation to $m=0$ of those of the $m$-walkers 
problem, since the 
eigenvalues do not necessarily have the same
analytic properties as the partition functions evaluated at fixed $t$.
\end{references}
\end{document}